\documentclass[dvips,12pt,a4paper]{article}
\usepackage{epsfig}
\usepackage{a4wide}
\usepackage{amssymb}
\begin{document}
\pagestyle{empty}
\vspace{1. truecm}
\begin{center}
\begin{Large}

{\bf The description of pp-interactions\\
with very high multiplicity at 70 GeV/c\\
by Two Stage Gluon Model}
\footnote
{Talk given at 7th International School - Seminar the Actual Problems
of Microworld Physics, Belarus, Gomel, 28.7-8.08,  2003}
\end{Large}

\vspace{2.0cm}
{\large  E. Kokoulina$^1$ and V. Nikitin$^2$}
\\[0.3cm]
$^1$ Gomel State Technical University, Belarus,
\begin{small}
{$helen_ {-}by@yahoo.com$}
\end{small}\\
\small\sl $^2$ Joint Institute for Nuclear Research, Dubna, Russia,\\
Vladimir.Nikitin@sunse.jinr.ru

\end{center}



\begin{abstract}
\noindent
Collective behaviours of secondary particles
in pp - interactions at 70 GeV/c are
researched. The Two Stage Gluon Model
is supposed for describing of processes
with very high multiplicity. It is shown
that gluons play the active role in the
multiparticle dynamics. Analysis of multiplicity
distributions of charged, neutral particles and
total multiplicity gives the thermodynamic meaning of
these interactions. The mechanism of the soft photon
formation  as a signature of the quark-gluon system
considered.
\vspace*{3.0mm}

\noindent
\end{abstract}

\section{Introduction}

These investigations were implemented in frameworks of
the project "Termalization". This project directs to
researches of collective behaviours of secondary
particles in proton-proton interactions at 70 GeV/c
\cite{SIS}.

On the basis of the present understanding of hadronic
physics, based on quantum chromodynamics~(QCD)~\cite{QCD},
as the theory of strong interactions, protons consist
from quarks and gluons. We construct the two stage
gluon model for the description of high energy
multiparticle production (MP) in proton interactions.
At first stage QCD and thermodynamical approaches
are used. At second stage (stage of hadronization)
the phenomenological description is applied \cite{TSM}.

After an inelastic collision of two protons the part
of the energy of these motive particles are converted
into the thermal one. Constituents of proton quarks
and gluons will have large energies, and they
can be described by perturbative QCD,
because the strong coupling reaches a small value.
Quarks and gluons become asymptotic free ones.
Our model investigations had shown that quark
division of initial protons in pp interactions
at 70 GeV/c is absent. Multiparticle production
is realized by active gluons.
These researches had confirmed the idea
P.Carruthers about a passive role
of quarks. He suggested that they are only "labels
and sources of colour perturbation in the vacuum:
meanwhile the gluons dominates in collisions and
multiparticle production "~\cite{CAR}.The domination
of gluons was first proposed by S. ~Pokorski and
L.~Van ~Hove \cite{PVH}. It was also noted that the
basis for active role of gluons is induced by the
existence of three-gluon or four-gluon couplings,
and this leads to the cascade of many gluons,
which are hadronized \cite{ALB}.

One of the most generally accepted methods
for study of multiparticle dynamics is the
multiplicity distributions (MD) analysis.
Using this method it is possible to known
much new about the development of study
processes. For this purpose two schemes are
supposed. They are distinguished only by
quark-gluon (QG) stage. If we want to study
gluon division inside QG system (QGS) we use
the first scheme (branch model). If we don't
interest what is going inside of QGS we use
second scheme (thermodynamic model). In both
schemes some of gluons (not of all) leave
QGS and convert to real hadrons or hadron
resonances. We named such gluons active
ones. In the thermodynamic idea we can
say that active gluons are evaporated
from hot QGS. After the evaporation
they pass stage of hadronization.

Basic research results were obtained
into frameworks of two built schemes
for MD of charged and neutral mesons,
total multiplicity. The dependence
of mean multiplicity of neutral pions
from number of charged particles was
obtained too. Compared parameters of
hadronization for charged and neutral
mesons we get the confirmation to the
hypothesis of the equal probabilities
of presence of quark-antiquark germs
of any flower into physical vacuum.
Usually this is known as a priori
probability.

Processes of pp interactions at 70~ GeV/c
were investigated experimentally some
years ago \cite{BAB}. MD of charged
particles were obtained. These distributions
were limited to 20 secondaries. Among them
were $n$ charged mesons ( $\pi^+$ or $\pi^-$)
and two leading protons:

\begin{equation}
\label{1}
p+p\rightarrow n\pi +2 N.
\end{equation}

Usually scaling variable $z=n/ \overline n$
is used, where $\overline n$ is the mean
multiplicity. For $n=20$ at 70 Gev/c we
have $z=3.5$. The kinematical limit constitutes
to $n_{\pi} =69$, where $n_{\pi}$ is the
number of pions. In present experiment it is planed
to get events with multiplicities $n_{\pi } =40 - 60$,
$z=5 - 7$. At very high multiplicities (VHMP)
and near the threshold of reaction (\ref{1})
all particles have a small relative momentum
and than the large density of hadron system
will be reached in the phase space. According
to generally accepted conception at these
conditions that system should be in QG plasma
(QGP) \cite{ENK} state.

The possible signal for QGP the most commonly
proposed ones being prompt production of
photons, prompt means that the photons should
not be decay products of hadron \cite{SOF}.
The explanation of the experimental increase
of the rate of direct soft $\gamma $'s as
the results of the Compton QG scattering
in comparison with calculations
based on the bremsstrahlung and
radiative decays of hadrons is proposed.
These photons could give the information
about early stage of QG interactions.

We are doing some suggestions on future
investigations. These schemes may be applied
for the description pp-interactions at the
higher energies (102, 205, 303, 405 and 800
GeV/c), others initial particles ($\pi^{+}$,
$\pi^{-}$, $K^{+}$, $K^{-}$, $\overline p$
and so on) and nucleus, if we modify our
models.

\section{Basing for  choice of scheme}
The choice of MP scheme is based on comparison
with experimental data~\cite{BAB}. At the
beginning of 90th successful description of
MD was realized by quark model \cite{CHI}.
In this model was suggested that one proton
quark pair, two pairs or three can to collide
and fragment into hadron jets. MD in quark jets
were described by Poisson. Second correlation
moments of charged particles for MD in this
model will be negative always. It is known
they are become positive at higher energies.
In addition this model doesn't take into account
situation when quark pair doesn't scatter but
secondary particles are appearing. In this model
it wasn't mentioned about gluons too.

Physicists from IHEP (Protvino) used generator
PYTHIA, as most applied for many purposes, and
obtained MD of charged hadrons \cite{BAB}.
They had shown that PYTHIA don't agree
with experimental data at high multiplicities
and has the deviation at $z=3.5$ equal to two orders.

In that way we must build new scheme of
hadron interactions for MD description. We want
to get agreement for very high multiplicity
\cite{SIS}(VHMP) region too. We use the Two Stage Model
(TSM) \cite{TSMA} which described MD in
$e^+e^-$-annihilation into hadrons from 10 to 200 GeV
well. We consider that at the early stage of pp
interactions initial quarks and gluons take part
in the formation of QGS. They can give branch
processes. On this stage (we name her of branch
stage) MD for quark and gluon may be described
Polya and Farry distributions \cite{GIO},
accordingly. On the second stage (hadronization
of quarks and gluons to hadrons) we take binomial
distributions \cite{TSM}.

Also as in TSM we use hypothesis of soft
colourless for quarks and gluons on the
second stage and add stage of
hadronization to branch stage by means
of factorization

\begin{equation}
\label{2}
P_n(s)=\sum\limits_{m=0} P_m^P(s)P_n^H(m,s),
\end{equation}
where $P_n(s)$ - resulting MD of hadrons,
$P_m^P$ - MD of partons (quarks and gluons),
$P_n^H(m,s)$~- MD of hadrons (second stage)
from $m$ partons. Generation function (GF)
for MD in hadron interactions are determined
by convolution of two stages

\begin{equation}
\label{3}
Q(s,z)=\sum\limits_{m=0}P_m^P(s)\left(Q^H(z)\right)^m=
Q^P(s,Q^H(z)),
\end{equation}
$$
P_n(s)=\frac{1}{n!}
\frac{\partial^n}{\partial z^n}\left
(Q^P(s,Q^H(z))\right)\biggl|_{z=0}\biggr.,
$$
where $Q^H$ and $Q^P$ - GF for MD at hadronization
stage and in QGS.

At the beginning researches we took model
where some of quarks and gluons from protons
take part in the formation of hadron jets.
Parameters of that model had values which
were differend a lot from parameters
obtained in $e^+e^-$- annihilation,
especially parameters of hadronization.
It was one of the main cause for refusal
from a scheme with active quarks.

After that we stopped on the model where
quarks of protons didn't take part in the
creation of jets, but remained into leading
particles. All of new hadrons were formed
by gluons. We will name these gluons
active ones. They could give branch
before hadronization. Part of these gluons
didn't convert into hadrons and they were
of source for  soft photons.

Other processes of scattering differ of
initial quark content. We think that if
at the beginning interaction antiquarks are
present side by side with quarks they can
as usually annihilate and make additional
leading mesons. That behaviour may be observed
in proton-antiproton scattering~\cite{RUS}.

It's very important to know how much active
gluons are into QGS at the first time after
the the impact of protons. We can assume that their
number may grow from zero and higher. It is
analogue of aimed parameter for nucleus.
Only in the case of elastic scattering
active gluons are absent. The simplest MD
for the  description of active gluons
formed in the moment of impact is Poisson
distribution

\begin{equation}
\label{4}
P_k=\frac{ e^{-\overline k} \overline k^k}{k!},
\end{equation}
where $k$ and $\overline k$ are the number and
mean multiplicities of active gluons, accordingly.
\section{TSM with branch}
We begin our MD analysis with branch scheme of
gluons. At the basing of scheme for description
of MP it was marked that on the first stage
in the moment of impact some active gluons may
appear. The energy of colliding protons is
transforming into internal energy of QGS. The
temperature of this system is raised sharply.
According to QCD gluons of protons may become
nearly free particles. We use MD (\ref{4}) for
the description of them. These active gluons
in QGS have some energy. If their energy is
large they may to give branch processes.
For the description of MD in gluon jets that
formed by branch process of $k$ active gluons
we use Farry distribution \cite{GIO}

\begin{equation}
\label{5}
P_m^B(s)=
\frac{1}{\overline m^k}\left(1-\frac{1}{\overline m}
\right)^{m-k}\cdot \frac {(m-1)(m-2)\cdots (m-k+1)}
{(k-1)!},
\end{equation}

if $k>1$ and
\begin{equation}
\label{6}
P_m^B(s)=\frac {1}{\overline m }\left (1-\frac {1}{\overline m}\right)^{m-1},
\end{equation}
$  if $\quad$
 k=1$,
where m and $\overline m$ are the number of secondaries
gluons and mean multiplicites of them. Expressions
(\ref{5})-(\ref{6}) were obtained from GF of one gluon
$Q_1^B$

\begin{equation}
\label{7}
Q_1^B=
\frac{z}{\overline m}\left[1-z \left(1-\frac{1}{\overline m}
\right)\right]^{-1}
\end{equation}
and from assumption about the independent branch of
gluons from each other, so GF for MD (\ref{5})-
(\ref{6})

\begin{equation}
\label{8}
Q_k^B=
\frac{z^k}{\overline m^k}\left[1-z \left(1-\frac{1}{\overline m}
\right)\right]^{-k}
\end{equation}
Mean multiplicity $\overline m$ is averaged to all
gluon jets.

We should mark the case at which $k=0$ (the impact
was elastic and active gluons are absent) resulted MD
of hadrons in pp-scattering is equal
$P_2(s)=e^{-\overline k}$.

On the second stage some of active gluons
may leave QGS and transform to real hadrons.We named
that gluons evaporated ones. Let us introduce parameter
$\alpha $ as the ratio of evaporated gluons, leaving QGS,
to all active gluons, which may transform to hadrons.
Our binomial distributions for MD of hadrons from evaporated
gluons on the stage of hadronization are

\begin{equation}
\label{9}
P_n^H=
C^{n-2}_{\alpha mN}\left(\frac{\overline n^h}
{N}\right)^{n-2}\left(1-\frac{\overline n^h}
{N}\right)^{\alpha mN-(n-2)}.
\end{equation}
$\overline n^h$ and $N$ are parameters in this
expression. They have the meaning average and
maximal possible multiplicity of hadrons from
one active gluon on the second stage. In this
expression an effect of two leading protons
is taking into account, too. GF for MD (\ref{9})
has form

\begin{equation}
\label{10}
Q_m^H=
\left (Q_1^H \right )^{\alpha m}=
\left [1-\frac{\overline n^h}{N}\left ( 1-z \right )
\right ]^{\alpha mN},
\end{equation}
where $Q^H_1$\quad- GF \quad for MD of one gluon,
\quad $\alpha m$ \quad -  the number of evaporated gluons
and $Q^H_1=\left [1-\frac{\overline n^h}{N}(1-z)
\right ]^{N}$.

MD of gluons on the first stage to finish
of branch may be written as

$$
P_m^P(s)=\sum\limits_{k=0}^{MK}\frac{e^
{-\overline k} \overline k^k}{k!}\sum\limits_{m=k}^{MG}
\frac {1}{\overline m^k}\left(1-\frac{1}{
\overline m}\right)^{m-k}
$$

\begin{equation}
\label{11}
\cdot \frac {(m-1)(m-2)\dots(m-k+1)}{(k-1)!}.
\end{equation}
Introducing in (\ref{2}) expressions (\ref{9})
and (\ref{11}) we obtain MD of hadrons in the
process of proton-proton scattering in two stage
gluon model (TSGM)

$$
P_n(s)=
\sum\limits_{k=0}^{MK}\frac{e^
{-\overline k} \overline k^k}{k!}\sum\limits_{m=k}^{MG}
\frac {1}{\overline m^k}
\frac {(m-1)(m-2)\dots(m-k+1)}{(k-1)!}
$$

\begin{equation}
\label{12}
\cdot \left(1-\frac{1}{
\overline m}\right)^{m-k}
C^{n-2}_{\alpha mN}\left(\frac{\overline n^h}
{N}\right)^{n-2}\left(1-\frac{\overline n^h}
{N}\right)^{\alpha mN-(n-2)}.
\end{equation}
Particular cases (an absent active gluons or
one gluon) don't mark in (\ref{12}). In
comparison with experimental data \cite{AMM}
the normalized factor $\Omega $ was introduced
to (\ref{12}). Numbers of gluons in sums on $k$
and $m$ were restricted by values $MK$ and $MG$
as maximal possible number of gluons on each stage.
We take for comparison data \cite{AMM} at $69$ GeV/c
because they are not differ from data at $70$ GeV/c
\cite{BAB}. $\chi^2$ in both cases are well about
$\sim 1$ at $70$ GeV/c and $\sim 10$ at $69$ GeV/c
and parameters are similarly.
Values of them  from comparison are

$N=40$ and more, \quad $\overline m=2.61 \pm .08$, \quad $\alpha =
.472 \pm .01$,

$\overline k=2.53 \pm .05$, \qquad $\overline n^h=2.50 \pm .29$, \qquad $\Omega =
4.89 \pm .10$ \\
with $\chi^2\approx 10$.
Values of $MK=6$ and $MG=1$ give the best $\chi^2$. So
we can conclude that branch processes are
absent at this energy and in this QGS. The part
of evaporating gluons is equal to 47.2 per cent.
Mean multiplicity of hadrons from one active
gluon equal $2.5$ and maximal possible number
of hadrons from gluon equal or more $40$. This
number looks very much like the number of
partons in the glob of cold QG plasma L.Van Hove
\cite{LVH}. If we are fixing parameter of hadronization
$\overline n^h$ and take it equal to$1.63$ as it will be
obtained in the thermodynamic model further,
our parameters will be have values

$N=40$ and more, \quad $\overline m=2.36 \pm .10$, \quad $\alpha =
.728 \pm .010$,

$\overline k=2.51 \pm .06$, \qquad $\overline n^h=1.63$ (fix.),
\qquad $\Omega =
2.15 \pm .18$ \\
with $\chi^2=2.9$ (see figure 1). At this case we have the part of
evaporating gluons is about 73 per cent, others
parameters are remained without considerable changes.
Effect of the evaporation of part active gluons may describe
appearance of soft photons. We will analyze this
effect further.

\section{Thermodynamic model}
In the thermodynamic model without branches appeared
in the moment of the impact active gluons may leave QGS
and fragment  to hadron jets. We consider that evaporated
from QGS active gluons have Poisson MD as (\ref{4})

$$
P_m=\frac{ e^{-\overline m} \overline m^m}{m!},
$$
and $\overline m$ is the mean multiplicity of that
gluons. Using binomial distribution for hadrons from
gluons (\ref{9}) and that idea convolution of two stages
(\ref{2}) we obtain MD of hadrons in pp-collisions in
framework two stage thermodynamic model (TSTM)

\begin{equation}
\label{13}
P_n(s)=\sum\limits_{m=0}^{ME}\frac{e^{-\overline m}
\overline m^m}{m!}
C^{n-2}_{mN}\left(\frac{\overline n^h}
{N}\right)^{n-2}\left(1-\frac{\overline n^h}
{N}\right)^{mN-(n-2)} (n>2)
\end{equation}
($P_2(s)=e^{-\overline m}$).
Our comparison (\ref{13}) with experimental data
\cite{BAB,AMM}\quad (see figure 2, the
point with n=2 was excepted)
gives that values of parameters

$N=4.24 \pm .13$, \quad $\overline m=2.48 \pm .20$, \quad
$\overline n^h=1.63\pm .12$\\
and normalized factor $\Omega =2$ with $\chi ^2\sim 2$.
We are constrained in sum~(\ref{13}) $ME=6$ (the maximal
possible number of evaporated gluons from QGS).
The significance of hadronization parameter $N$ at
the description of experimental data of $e^+e^-$
annihilation was fined equal to $\sim 4-5$ \cite{TSM} .
We can see that our parameter $N$ obtained in TSMT
coincides with this value. Besides both models TSMB and
TSTM describe data well.

From TSTM the maximal possible of number charged
particles is $26$. This quantity  is the product
of maximal multiplicities active gluons and
maximal number of hadrons formed from one gluon
$ME \cdot N$. In TSMB we have more hadrons, but
with very small probabilities.

It is interesting to get MD for neutral mesons.
For this purpose we will take experimental
mean multiplicity of $\pi ^0$'s in pp-intaractions
at 69 GeV/c. It is equal to $2.57\pm .13$ \cite{MUR}.
Since the mean multiplicity in this process is
calculated as the product of mean number of gluons
$\overline m$ and hadron parameter $\overline n^h$
we can determine parameter of hadronization
$\overline n^h_0$ for neutral mesons. It's equal
to $1.036$. We do the simplification on the second
stage as in TSM \cite{TSM} for different particles
(equalities of probabilities of the
creation of different hadrons)

\begin{equation}
\label{14}
\frac {\overline n^h_{ch}}{N_{ch}}\approx
\frac {\overline n^h_0}{N_0}\approx
\frac {\overline n^h_+}{N_+}\approx
\frac {\overline n^h_-}{N_-}\approx
\frac {\overline n^h_{tot}}{N_{tot}},
\end{equation}
were lower indexes $ch$, $0$, $+$, $-$  and $tot$ determine
what this quantity corresponds to a charged particle,
$\pi ^0$, $\pi ^+$, $\pi ^-$ and  for total multiplicity
$n_{tot}=n_{ch}+n_0$, accordingly. Parameters of hadronization $N$ for
neutral, positive, negative and total multiplicity
are determined from knowing according $\overline n^h$'s
and $N_{ch}$. MD for neutral mesons  $\pi ^0$'s,
($\pi ^{+}$'s, $\pi ^-$'s, total multiplicity) have
form (\ref{13}) and may be easy obtain if they
will be normalized to mean multiplicity
$\pi ^0$'s ($\pi ^{+}$'s, $\pi ^-$'s, total multiplicity).
The mean multiplicities of $\pi ^+$ or $\pi ^-$
mesons are determined as $(\overline n_{ch}(s)-2)/2$,
the total mean multiplicity equal to the sum of
charged and neutral mean multiplicities:
$\overline n_{tot}(s)=\overline n_{ch}(s)+
\overline n_{0}(s)$.
The expression (\ref{12}) gives analogous results, but
we use more parameters). MD for neutral mesons are giving
on figure 3. From this distribution we see that
the maximal possible number of $\pi ^0$'s
from TSTM is equal 16. MD for total multiplicity are giving
on figure 4. We see that the maximal possible number of total
particles in this case is equal 42. In TSMB the maximal
possible number of particles is more but probabilities for
them are very small. In TSMB we can study what has happened
inside of QGS and outside of it (the evaporation of some
active gluons). We should note that data were obtained
at $n_{ch} < 20$. In the project Thermalization they want
to get events with very high multiplicities
$n_{ch} > 20$. It's possible that branch will begin
at that ones.

In the conclusion of this section the dependence of the
mean multiplicities of neutral mesons versus the number
of charged particles will be obtained. With the help
of MD for total multiplicity $P_{n_{tot}}(s)$ and take
into account of Bayess theorem we have

\begin{equation}
\label{15}
\overline n_0(n_{ch},s)=\frac {\sum \limits _{n=n_1}^{n_2}
P_{n_{tot}}(s) \cdot (n-n_{ch})}{\sum \limits _{n=n_1}^{n_2}
P_{n_{tot}}(s)},
\end{equation}
where $\overline n_0(n_{ch},s)$ - the mean multiplicity
of $\pi _0$'s, $n_{ch}$ - the number of charged
particles, $P_{n_{tot}}(s)$ - MD for total multiplicity,
$n_1$ and $n_2$ are lower and top boundaries for total
multiplicity at given number of charged particles
$n_{ch}$. Obtained by TSTM (figures 2-4) MD of charged
and neutral secondaries give maximal number for charged
$n_{ch}=26$, neutral $n_{0}=16$ and total $n_{tot}=42$.
There fore we have next limits to $n=n_1$ and $n=n_2$:
$n_1 \geq n_{ch}$,  $n_2 \leq 16+ n_{ch}$. These
restrictions are determined only by conservation laws.
Mean multiplicity for these limits has form as
figure 5. We can see the big distinction with experimental
data \cite {GRI,PIO} at small multiplicities.

The marked improvement will be reached if we decrease
top limit at low multiplicities ($n_{ch}\leq 10$) to
$n_2=2 n_{ch}$. That is corresponding to the case, when
the maximal number of neutrals is equal to the number of charged,
and it is possible double excess neutral mesons over positive
(negative) pions.
At bigger charged multiplicites this
limit is defined of $n_2=16+n_{ch}$ (the maximal
possible number of neutrals which can be created).
We don't know what is happening at the region of VHMP
with $\overline n^0(n_{ch})$.
On the figure 6 it is shown that multiplicity of neutrals
versus $n_{ch}$ when $n_2$ is taken equal to $2 n_{ch}$
at small $n_{ch}$ and $n_2=16+n_{ch}$ at $n_{ch}>10$.
It should be marked that the value of the low limit $n_1$
is staying almost constant and equal to 0 ($n_{ch}<16$),
0-1 ($n_{ch}=16-18$) or 1-2 ($n_{ch}>18$). Such behaviour
of $n_1$ and $n_2$ in (\ref{15}) indicates that
Centauro events \cite{CEN} with a large charged particles and
practically no accompanying neutrals may be
realized in the region of VHMP.

AntiCentauro events with a large number of neutrals and
with very small charged must be absent. On the figure~
5 it is seeing that the existence of antiCentauro events
at small $n_{ch}$ must to give very large mean
multiplicity of neutrals that is contradicting experimental
data \cite{MUR}$^-$\cite{PIO}.

\section{Soft photons and hypothesis of
a priori probability}
In two stage model with gluon branch it was shown
that several of active gluons are staying inside of
hot QGS and don't give hadron jets. What has happened
with such gluons at the hadronization? New formed
hadrons don't contain of gluon content inside and
around themselves. They are catching up
small energetic gluons which were free before this
time. Gluons have possibility to stick to them.
Owing to the large strong coupling on this stage
they must adhere to just new formed hadrons
(the confinement of gluons). These hot hadrons
are excited because they have additional energy
at the expense of absorbed gluons. This energy
may be thrown down by means of the photon
radiation.

The production of photons in particle collisions
at high energies has been studied in many
experiments from 17 to 1800 GeV \cite{CHL}.
Project Themalization is planning to conduct
investigations of low energetic photons with
$p_t \leq 0.1GeV/c$ and $x \leq 0.01$
\cite{NIK}. Usually such photons are named
soft photons (SP). The experimental spectra of
SP were obtained. It was shown that measured
cross sections of such photons are several times
larger than expected from QED inner bremstruhlung.
For the explanation  the excess of SP phenomenological
models were suggested. The most known of them are
the glob model of Lichard and Van Hove \cite{LVH}
and the modified soft annihilation model Lichard
and Thomson \cite{SAM}.

We want to understand what is the souse of such
SP and to estimate the number of them. We consider
that in the certain moment of time QGS or exited
new hadrons may set in almost equilibrium state
on the short or finite time. That's why we will try
to use for the description of the massless bosons
(gluons and photons) the black body emission
spectrum \cite{HER}

\begin{equation}
\label{16}
\frac {d \rho (\nu )}{d \nu }=
\frac {8\pi }{c^3} \frac {\nu ^2}{e^{h\nu /T}-1},
\end{equation}
where $ \nu $ is the energy of photon. These result
spectra could help us  to calculate the
number of SP. It should note that the developing
in the real time process of hadron interaction
reminds Big Bump \cite{KUR} one resulting our
Universe creation. In the last case the equilibrium
between electron-proton and photons breaks at very
high temperature when electron-proton recombination
and the creation neutral atoms take place. At this
stage photons cease to interact with hadrons and
begin to expand as relict ones. At the present
time we observe the spectrum of relic photons
\cite{HER}.

The gluon density at the deconfinement
temperature $T_c\approx 160-200$ MeV can be
estimated by comparison with relic one

\begin{equation}
\label{17}
\rho _{gl} (T )=
5.479 \left(\frac {T}{T_0}\right)\cdot 10^{-37} (fm)^{-3},
\end{equation}
where $T_0\approx 3^o$K - the temperature of contemporary
relic photons and $\rho =5.479 \cdot 10^{-37}(fm)^{-3}$
is the density of them. Gluon densities at the
deconfinment temperature $T_c=160$ or $200MeV$ are
$\rho _{gl}(160)=0.13 (fm)^{-3}$ and
$\rho _{gl}(200)=0.25 (fm)^{-3}$. The number of
gluons $N_{gl}$ in the hot QGS of size $\sim L^3$,
where $L=20$fm, will be the order of thousands:
$$
N_{gl}(160)\sim 1000, \quad
N_{gl}(200)\sim 2000.
$$
In the case $L\sim 10-19$ fm $N_{gl}\sim 100-900$.
We conclude that inside our QGS has a many particles.
That system may be described by statistical and
thermodynamic methods.

Using the spectral spatial density of relic
photons (\ref{16}) again it is possible to
get the number of SP $N_\gamma $ in the region
of size of our system (new formed hadrons).
This size must be bigger than one in the
gluon case. Dependencies of multiplicities
of SP from the energy (the moment $p$) and
the linear size of system (L) are given in
Table 1.

\begin{center}
Table 1. Multiplicities of soft photons $N_{\gamma }$.
\end{center}
\renewcommand{\tablename}{Table}
\begin{center}

\begin{tabular}{|c|c|c|c|c|}
\hline
$ \qquad p, MeV/c \qquad $ &$\qquad  10 \qquad $ &$
\qquad 15 \qquad  $&$ \qquad  20 \qquad  $&$ \qquad 30 \qquad$  \\[1ex]
\hline
\hline
$L,fm$ &$N_{\gamma}$&$N_{\gamma}$&$N_{\gamma}$&$N_{\gamma}$\\[1ex]
\hline
$50$&$3.96$&$13$&$32$&$107$\\[1ex]
\hline
75&$13.36$&$45$&$107$&$361$\\[1ex]
\hline
100&$31.68$&$107$&$253$&$855$\\[1ex]
\hline
$120$&$64.87$&$209$&$495$&$1670$\\[1ex]
\hline
\end{tabular}
\end{center}

If the size of our system about 50 fm
and average energy of photons 15-20~MeV/c
the number of such SP will be the order
of 20.

In the conclusion of this section we will
analyse the hypotheses of a priori
probabilities. It's assumed that the beginnings
different quark pairs from the physical vacuum
happen with equal probabilities. We will try
to examine it by the example of formation
neutral and charge mesons. On the whole at this
energy $u \overline u$ and $d \overline d$ quark
pairs may appear. If process of hadronization
is begun from the formation one charged meson than
the opposite charged meson must form on the force of
the low of charge conservation. In the contrary case
at the formation of neutral it's not necessary
the creation of additional particles. So we can
affirm that the number of charged hadrons will be more
than neutral ones, or the probability of the creation of
charged more than neutral. We can give simple estimate
of these probabilities.

Two parameters of
hadronization were obtained by TSTM $\overline n_{ch}$
and $\overline n_0$. In according to this hypothesis
(a priory  probabilities) MD of $\pi _0$'s from one
gluon jet on the second stage may be described
by the binomial distribution

\begin{equation}
\label{18}
P_{n_0}=
C_{N_t}^{n_0}(0.5+\delta )^{n_0}(0.5-\delta )^{N_t-n_0},
\end{equation}
here $N_t$ - the maximal possible number of mesons
formed from gluon,
$n_0$ - the number of neutral mesons among these
secondaries (the number of charge mesons
$n_c=N_t-n_0$). The probability of creation
one pair of charge particles ($\pi ^+\pi ^-$)
is $p_c=0.5+\delta $ and the probability for
one neutral pion is $p_0=0.5-\delta $. The normalized
condition is $p_0+p_c=1$. From TSTM it was obtained
$\overline n_{ch}=1.63$ and $\overline n_{0}=1.036$.
In accordance with (\ref{17}) these mean multiplicities
for binomial distributions will be equal to next
expressions
$$
\overline n_{ch}=(0.5+\delta )N_t,
$$
\begin{equation}
\label{19}
\qquad \overline n_{0} =(0.5-\delta )N_t.
\end{equation}
The probability of the creation of charge pair
is more than the neutral meson one ($\overline n_{ch} >
\overline n_{0}$). The decision of (\ref{19}) gives
$N=2.666$ and $\delta =0.247$. Than probabilities
of creation $p_0=.253$ and $p_{c}=0.747$. The ratio
of these values is $p_c/p_0\sim 3$.

\section{Conclusions}
The two stage gluon model for the description
of pp-interactions with VHMP at 70 GeV/c
was supposed. It's shown the important role of active
gluons in the formation of new hadrons. MD for charge
particles are described well. MD for neutral mesons
and total multiplicity are obtained too.
The maximal possible number of charge and
neutral mesons in these processes is found . The thermodynamic
idea of the gluon evaporation are given.

Sources of soft photons is researched, and the number
such $\gamma $'s is calculated by statistical physics
methods. The analysis of soft gluons are implemented
too. The mechanism of creation neutral and charge
mesons are investigated.It is ascertained that Centauro
events may be discovered in the region charged VHMP
and the existence of antiCentauro events is rejected.

\section*{Acknowledgments}
We would like to thank Sissakian A.N.
for support and encourage of our investigations,
Kuraev E.A. for the help in the
understanding of soft photons nature and others
of scientists from LTP JINR (Dubna, Russia).
One of us (Kokoulina E.S.) is deeply indebted to
chief of Laboratory of Physical
Investigations of Gomel Technical University
Pankov~ A.A. and to all
collegues from Gomel Technical
University for supporting her job many years.

\newpage

\begin{figure}
\begin{minipage}[b]{.3\linewidth}
\centering
\includegraphics[width=\linewidth, height=2in, angle=0]{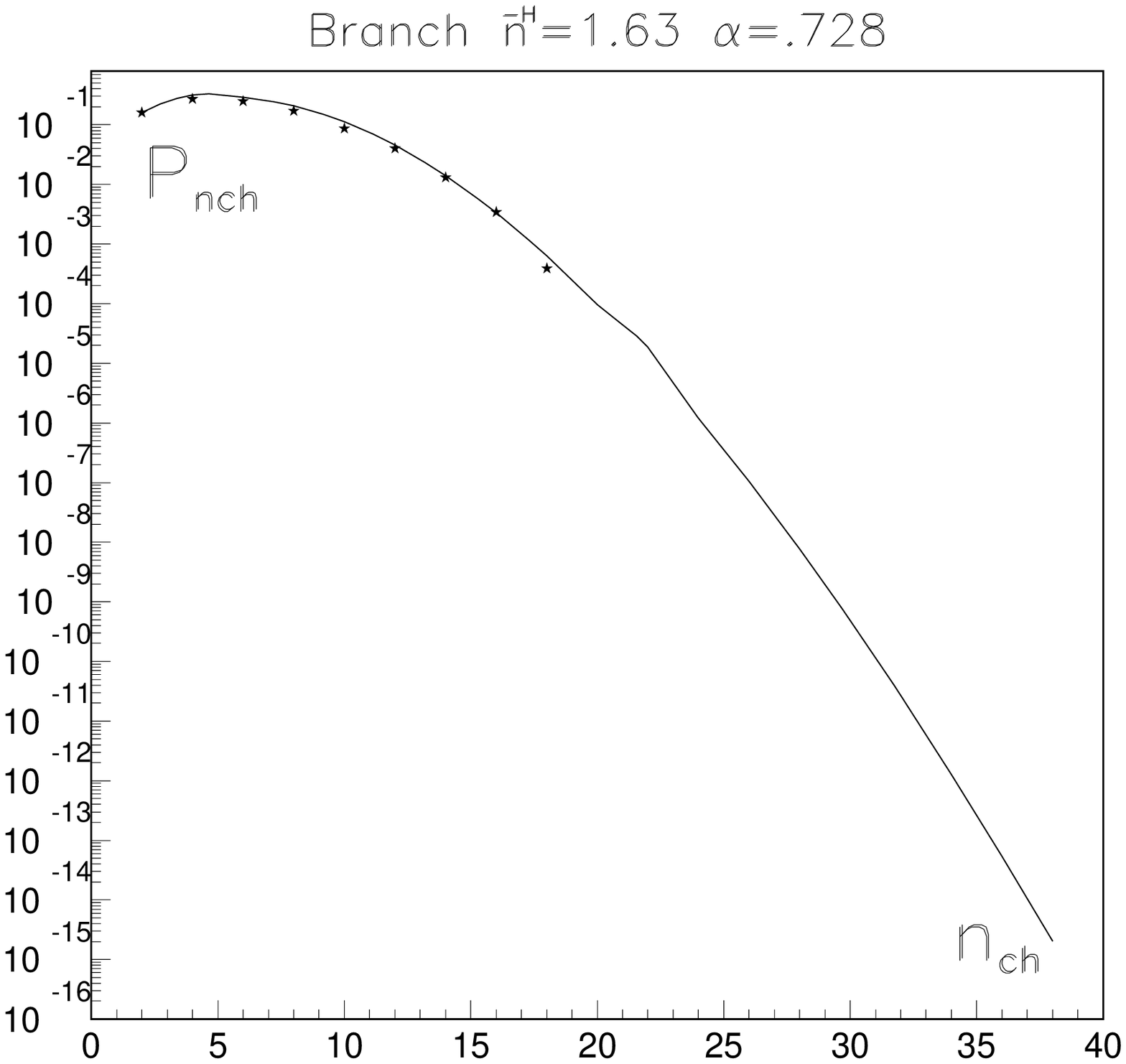}
\caption{MD $P(n_{ch})$ in TSMB.}
\label{1dfig}
\end{minipage}\hfill
\begin{minipage}[b]{.3\linewidth}
\centering
\includegraphics[width=\linewidth, height=2in, angle=0]{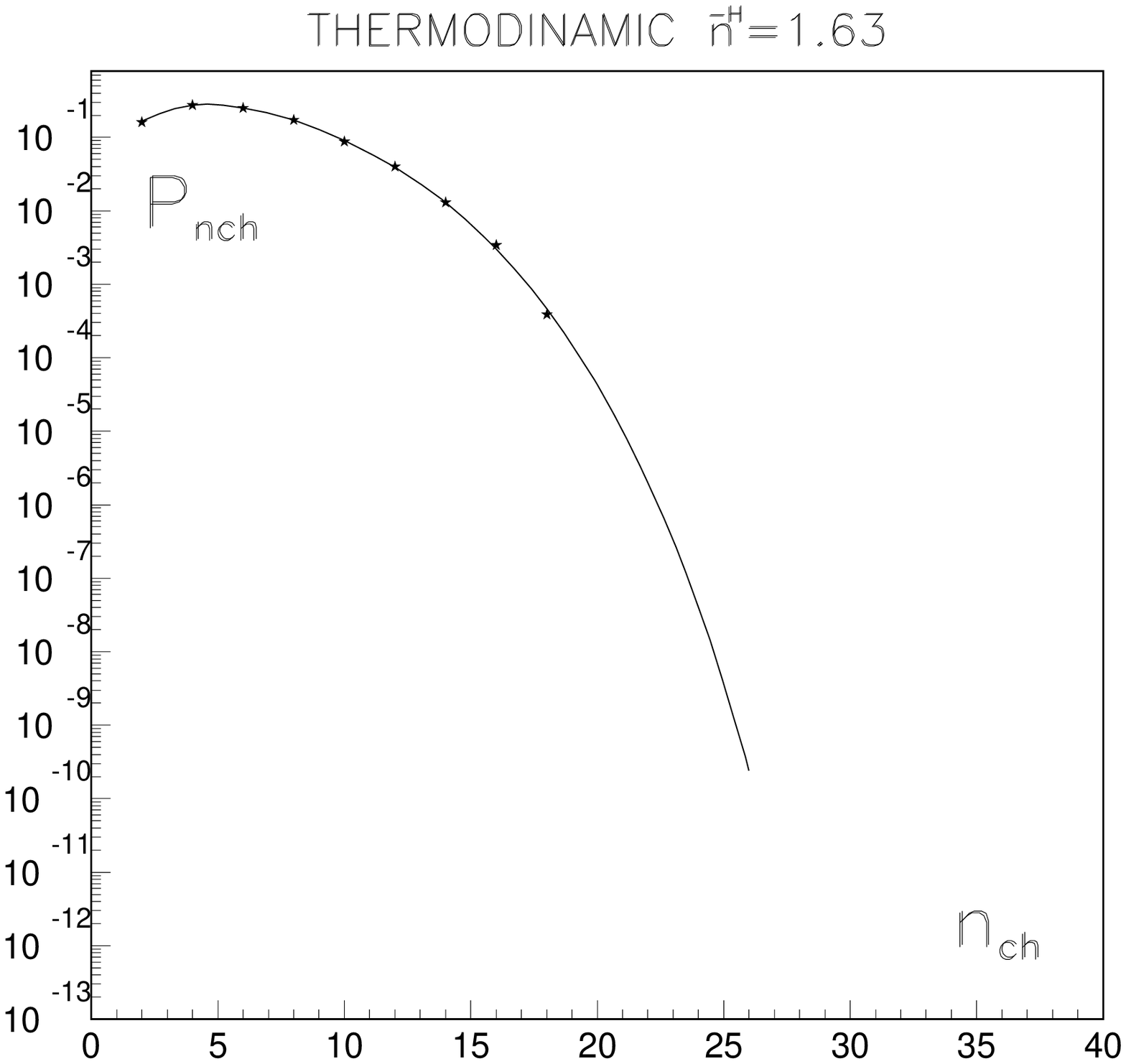}
\caption{MD $P(n_{ch})$ in TSTM.}
\label{2dfig}
\end{minipage}\hfill
\begin{minipage}[b]{.3\linewidth}
\centering
\includegraphics[width=\linewidth, height=2in, angle=0]{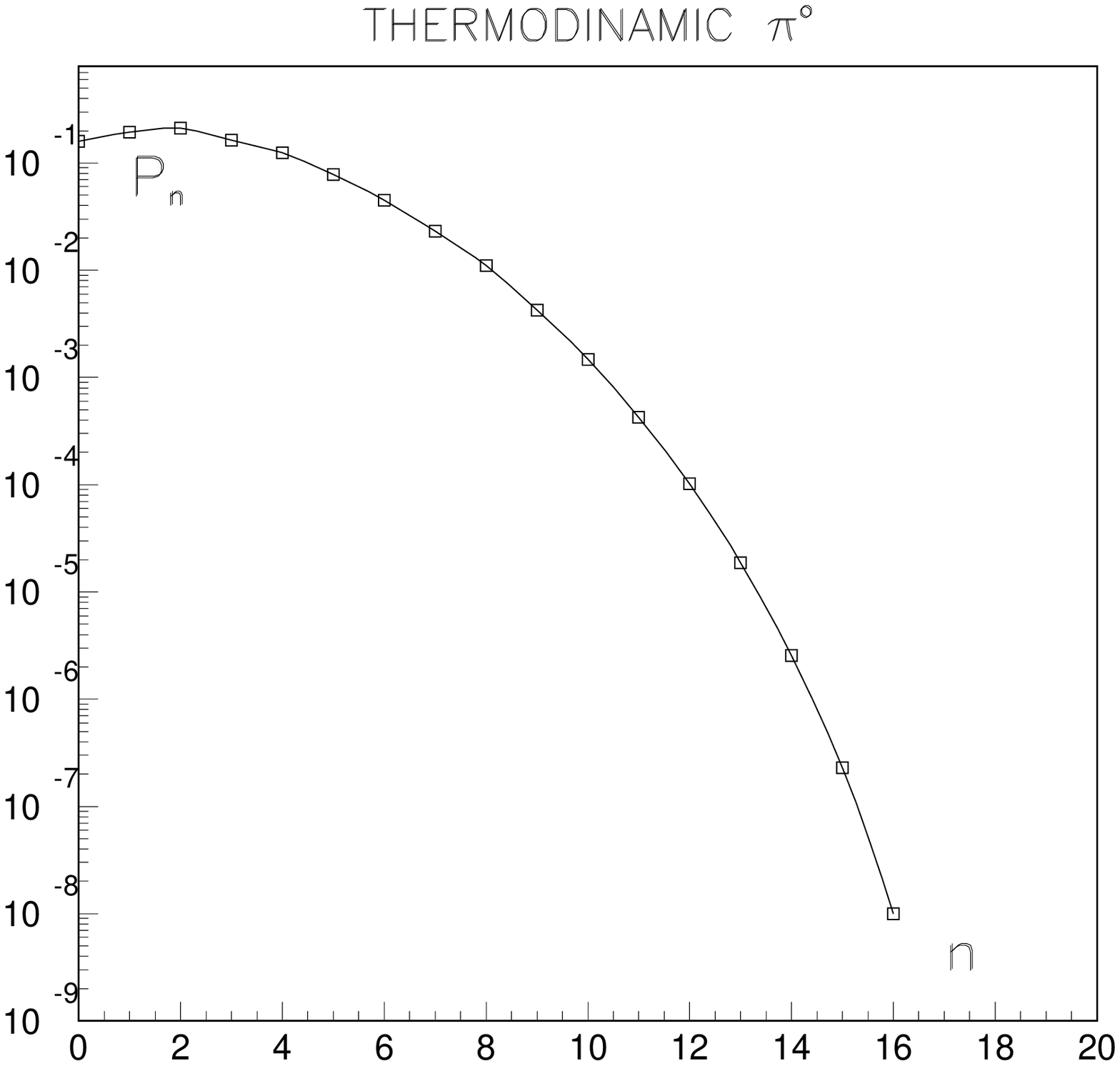}
\caption{MD $P(n_o)$ in TSTM.}
\label{3dfig}
\end{minipage}
\end{figure}

\begin{figure}
\begin{minipage}[b]{.3\linewidth}
\centering
\includegraphics[width=\linewidth, height=2in, angle=0]{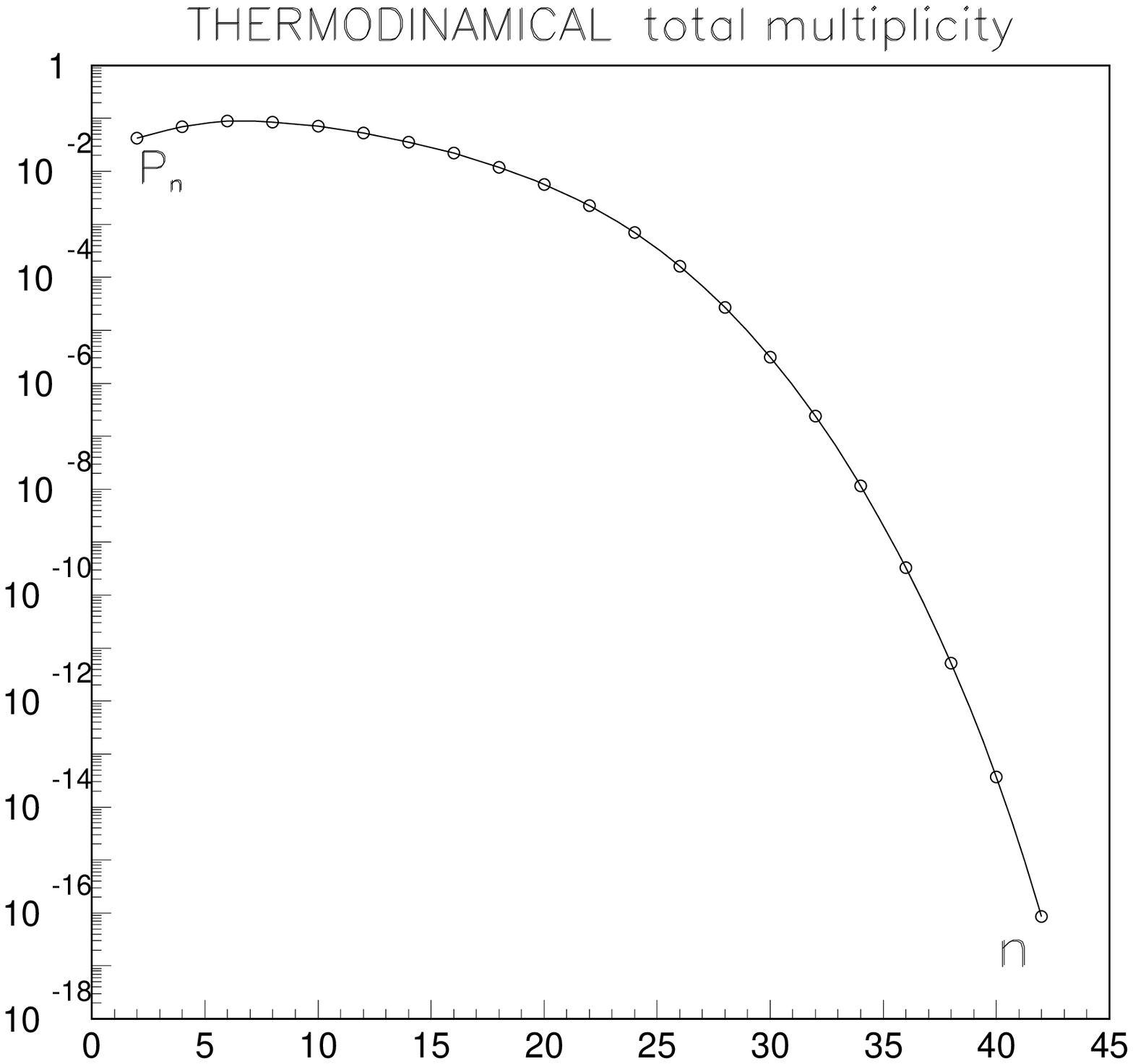}
\caption{MD $P(n_{tot})$ in TSTM}
\label{4dfig}
\end{minipage}\hfill
\begin{minipage}[b]{.3\linewidth}
\centering
\includegraphics[width=\linewidth, height=2in, angle=0]{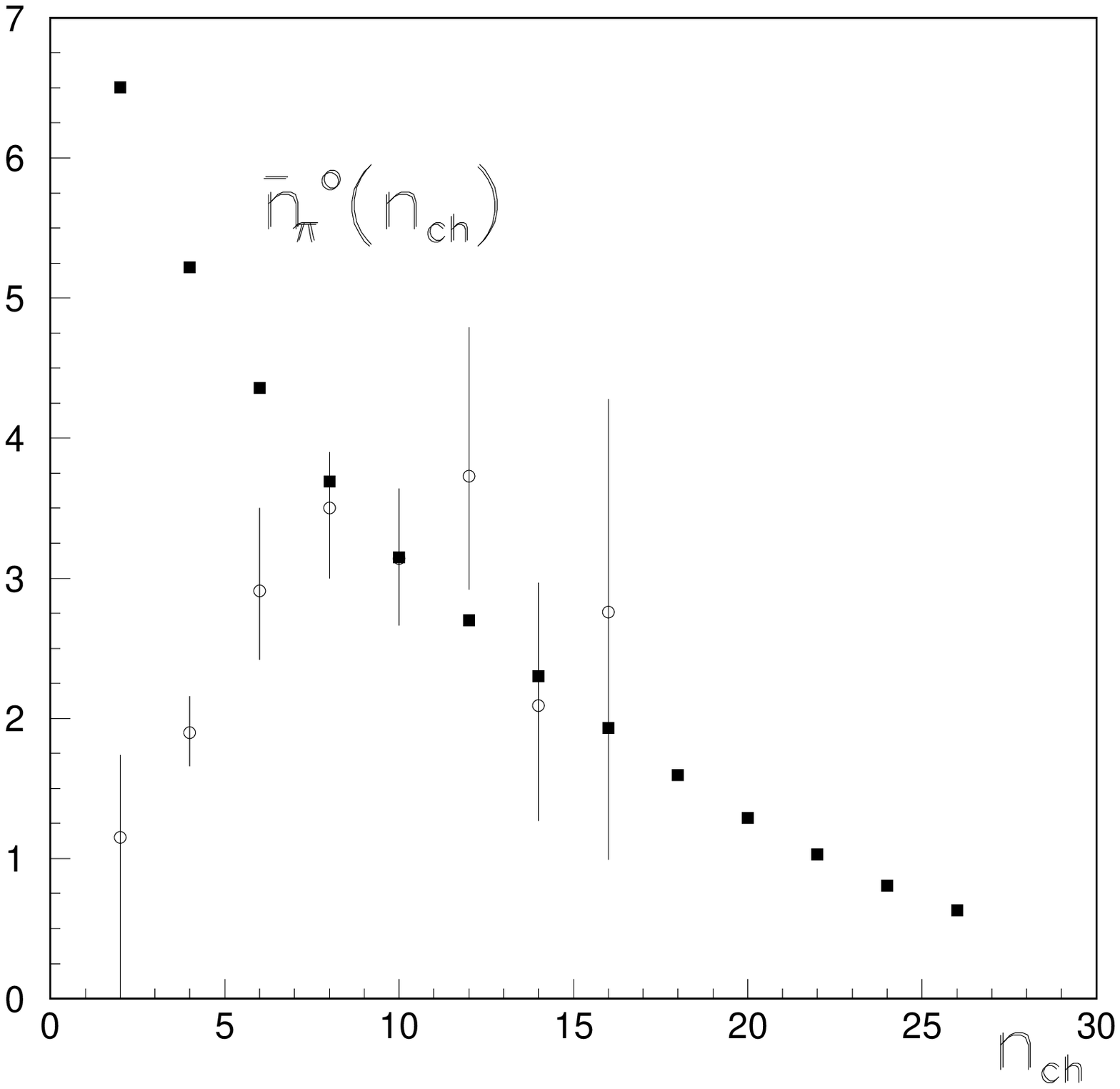}
\caption{$\overline n_\pi^o$ versus $n_{ch}$ (see text).}
\label{5dfig}
\end{minipage}\hfill
\begin{minipage}[b]{.3\linewidth}
\centering
\includegraphics[width=\linewidth, height=2in, angle=0]{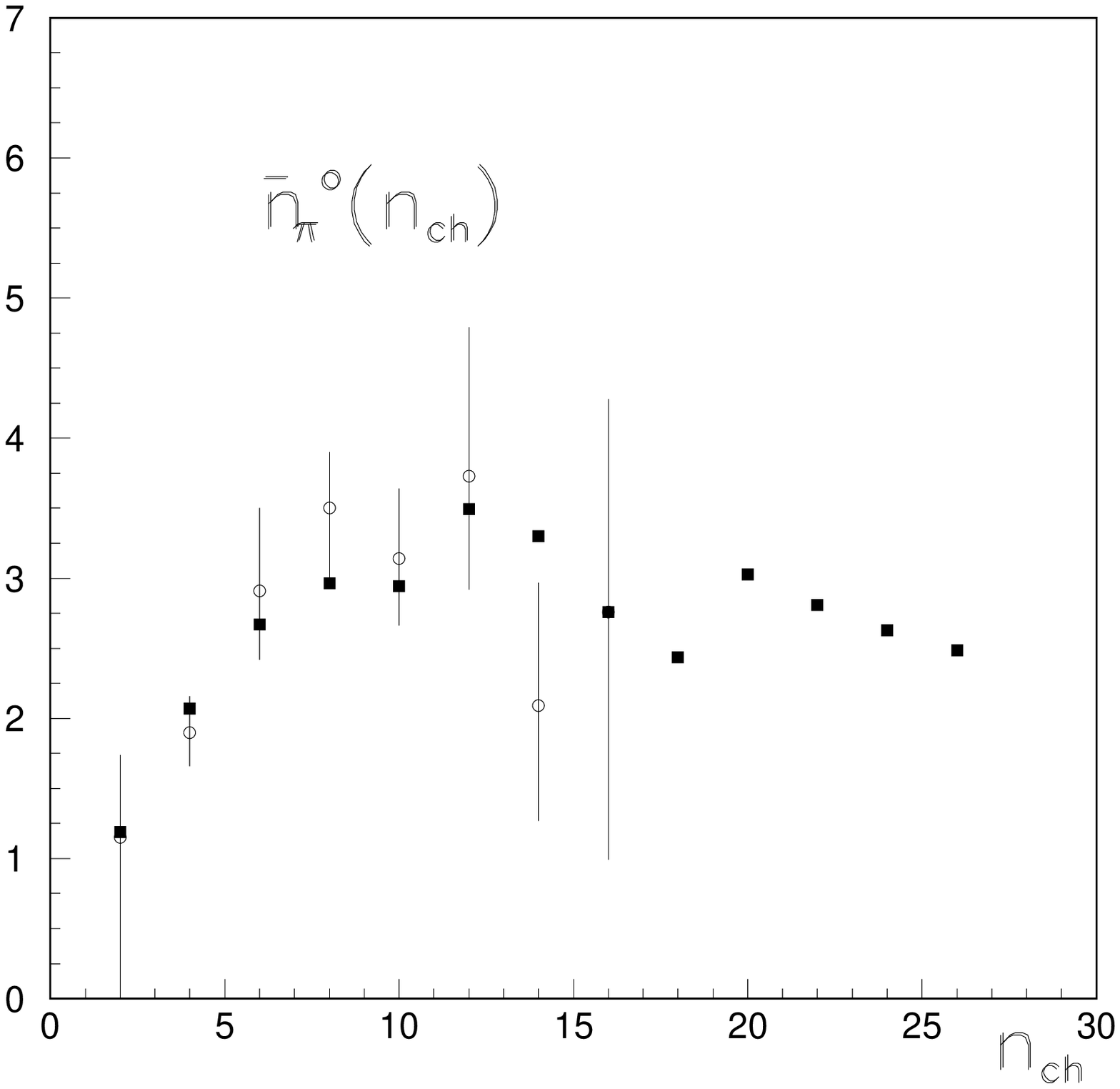}
\caption{$\overline n_\pi^o$ versus $n_{ch}$ (see text).}
\label{6dfig}
\end{minipage}
\end{figure}

\end{document}